# Curvature Effects in Special Relativity

Moninder Singh Modgil

*Department of Physics,*

*Indian Institute of Technology, Kanpur,*

*India*

Email: msingh@iitk.ac.in

Date: 16 August 2004

**Abstract**

Space-time measurements, of gedanken experiments of special relativity need modification in curved spaces-times. It is found that in a space-time with metric g, the special relativistic factor $\gamma$, has to be replaced by $\gamma_g = \dfrac{1}{\sqrt{g_{\mu\nu} V^\mu V^\nu}}$, where $V^\mu = (1,v,0,0)$, is the *4*-velocity, and *v* the relative velocity between the two frames Examples are given for Schwarzschild metric, Freidmann-Robertson-Walker metric, and the Gödel metric. Among the novelties are paradoxical tachyonic states, with $\gamma_g$ becoming imaginary, for velocities less than that of light, due to space-time curvature. Relativistic mass becomes a function of space-time curvature, $m = \sqrt{g_{\mu\nu} P^\mu P^\nu}$, where $P^\mu = (E,\boldsymbol{p})$ is the *4*-momentum, signaling a new form of Mach's principle, in which a global object – namely the metric tensor, is effecting interia.

# Curvature Effects in Special Relativity

## 1. Introduction

It is generally considered that the laws of physics can locally be described in a flat space-time setting. This is described by the statement that locally any space-time is Minkowskian. However, space-time curvature can significantly effect local physics. One example of this is the Hawking radiation [1]. Another example is the effect of acceleration on reading of particle detectors in curved space-times [2]. Space-time curvature effects local structure of light cone, and the behaviour of null geodesics. Since gedanken experiments of Special Theory of relativity (STR) are based upon propagation of light signals, therefore. it may be enquired, how would the various results of STR get modified, when the space-time is curved.

Special theory of relativity (STR) is based upon the Minkowski metric,

$$ds^2 = \eta_{\mu\nu} dx^\mu dx^\nu, \qquad \mu,\nu = 0,1,2,3 \tag{1}$$

In a curved space-time with the line element,

$$ds^2 = g_{\mu\nu} dx^\mu dx^\nu, \tag{2}$$

effect of local curvature, in the form of components of metric tensor, $g_{\mu\nu}$ may be expected to show up in the local measurements. The line element of a null geodesic in both flat and curved space-times is characterized by,

$$ds^2 = 0. \tag{3}$$

This equations encodes the fundamental principle of STR, namely the constancy of speed of light in all inertial frames. It turns out that this is the key equation, for deriving equations for STR in both flat as well as curved space-times – derived in this paper. In this paper we investigate this issue, and derive time dilation, length contraction, energy-mass relations for STR in a curved space-time setting. We use Schwarzschild metric, Robertson-Walker metric, and Gödel metric, as examples for this.

Now, consider the familiar gedanken experiment of special relativity in which a beam of light is sent from ceiling to floor of a train moving with velocity $v$, with respect to the ground. Let F be the frame moving along with train, and $F_{ground}$, be the frame fixed to ground. Let height of the train carriage be denoted by $h$. Let $x$ direction be horizontal along the ground, and $y$ direction, vertical, while $t$ stands for the time co-ordinate in $F_{ground}$. In F the time co-ordinate is represented by $t'$. The $z$ direction is being ignored for the present. In a time interval $dt$, in frame $F_{ground}$ the train travels a distance,

$$dx = vdt. \qquad (4)$$

In this much time interval, a ray of light, covers the distance $h$, from the ceiling to the floor of the train carriage,

$$dy = h. \qquad (5)$$

In the frame F moving along with train, time taken for this journey is $dt'$.

Let the metric be diagonal. However, the result derived below can remains unchanged, when the metric has an off diagonal element such as $g_{10}$. Equating line element of a null geodesic in the co-moving frame F, to zero, one gets,

$$ds^2 = 0 = g_{00}dt'^2 - g_{22}dy^2. \qquad (6)$$

Line element for null geodesic as seen in the ground fixed moving frame F $_{gruond,}$ is,

$$ds^2 = 0 = (g_{00} - g_{11}v^2)dt^2 - g_{22}dy^2 \qquad (7)$$

From above two equations it follows that,

$$dt' = \sqrt{\frac{g_{00} - g_{11}v^2}{g_{00}}}dt', \qquad (8)$$

which is the formula for special relativistic time dilation in a curved space-time. This may be re-written as,

$$dt' = \sqrt{\frac{g_{\mu\nu}V^\mu V^\nu}{g_{00}}}dt' \qquad (9)$$

where $V^\mu$ is the *4*-velocity,

$$V^\mu = (1, v, 0, 0). \qquad (10)$$

In a general relativistic frame work, using conformal time, $g_{00}$ can be equated to *1*, giving,

$$dt = \frac{dt'^2}{\sqrt{g_{\mu\nu}V^\mu V^\nu}}. \qquad (11)$$

Thus we can define a general relativistic $\gamma_g$,

$$\gamma_g = \frac{1}{\sqrt{g_{\mu\nu}V^\mu V^\nu}}, \qquad (12)$$

for special relativistic transformations in a curved space-time with metric *g*. We note that the usual,

$$\gamma = \frac{1}{\sqrt{1-v^2}},\qquad(13)$$

of Minkowski metric, can be expressed as,

$$\gamma = \frac{1}{\sqrt{\eta_{\mu\nu}V^\mu V^\nu}}.\qquad(14)$$

Next consider the gedanken experiment in which the ray of light is sent from one end of carriage to another in a horizontal direction. Let *dx'* be the length of carriage in F and *dx* the length in F $_{ground.}$. A mirror is mounted on the other end of carriage, which reflects the ray back. Time for round trip in F is *dt'*, while in F $_{ground}$ it is *dt*. Equating line elements of null geodesics in F and F $_{ground}$ to zero we have,

$$g_{00}dt'^2 - g_{11}dx'^2 = g_{00}dt^2 - g_{11}dx^2 = 0.\qquad(15)$$

Using expression eq.() for time dilation, this gives the required formula for length contraction,

$$dx' = \sqrt{\frac{g_{00}}{g_{\mu\nu}V^\mu V^\nu}}dx,\qquad(16)$$

which for $g_{00}=1$, may be re-written as,

$$dx' = \frac{dx}{\sqrt{g_{\mu\nu}V^\mu V^\nu}}.\qquad(17)$$

**2. Lorentz transformations**

Lorentz transformations in a curved space-time for a *4*-vector $X^\mu = (x^0, x^1, x^2, x^3)$, may be obtained from those for a flat space-time, by replacing $\gamma$ by $\gamma_g$,

$$x^{0'} = \gamma_g (x^0 - vx^1),$$

$$x^{1'} = \gamma_g (x^1 - vx^0),$$

$$x^{2'} = x^2,$$

$$x^{3'} = x^3. \tag{18}$$

Proper time $\tau$ in GTR is essentially defined by the equation of line element,

$$d\tau^2 = g_{\mu\nu} dx^\mu dx^\nu, \tag{19}$$

which obscures effect of relative velocity between inertial frames. However, using a *4*-velocity,

$$V^\mu = (v^0, v^1, v^2, v^3), \tag{20}$$

and putting co-ordinate differentials,

$$dx^i = v^i dt, \tag{21}$$

equation for proper time may be re-written as,

$$d\tau^2 = g_{\mu\nu} V^\mu V^\nu dt = \frac{dt^2}{\gamma_g^2}. \tag{22}$$

Proper velocity may be defined as,

$$\eta^\mu = \gamma_g v^\mu. \tag{23}$$

Effect of space-time curvature on relativistic mass $m_g$ can be calculated, using the usual collision experiment between two particles. Assuming the usual non-relativistic definition of momentum, as product of mass times velocity, along with law of conservation of (rest) mass, does not lead to, momentum conservation in the moving frame. Therefore the momentum is re-defined using proper velocity. Let subscripts $A$ and $B$ denote incoming particles, and $C$ and $D$ outgoing particles. Conservation of momentum between the stationary and the moving frames, gives,

$$m_A \eta_A + m_B \eta_B = m_C \eta_C + m_D \eta_D, \qquad (24)$$

$$m_A \eta_A^0 + m_B \eta_B^0 = m_C \eta_C^0 + m_D \eta_D^0, \qquad (25)$$

where, the superscript $0$ on $\eta$ indicates stationary frame. The law of conservation of momentum holds, provided the conserved quantity is the relativistic mass,

$$m_g = \gamma_g m_0, \qquad (26)$$

where $m_0$ is the rest mass. We have the following definitions for relativistic momentum and energy,

$$p^i = m_0 \eta^i = \gamma_g m_0 v^i, \qquad (27)$$

$$E = m_0 \eta^0 = \gamma_g m_0. \qquad (27)$$

Energy-mass relation would be given by,

$$m_g^2 = g_{\mu\nu} P^\mu P^\nu, \qquad (28)$$

where $P^\mu = (p^0, p^1, p^2, p^3)$, is the 4-momentum, and $p^0 = E$, is energy as usual. Above equation indicates that relativistic mass is a function of local curvature, which is a new form of Mach's principles [3], in the sense that inertia is now a function of local value of a global object, namely the space-time curvature tensor. Note that energy-mass expression in STR,

$$E^2 - p^2 = m^2, \qquad (29)$$

can be written as,

$$m^2 = \eta_{\mu\nu} P^\mu P^\nu. \tag{30}$$

## 3. Some examples

For Schwarszchild line element [4],

$$ds^2 = \left(1 - \frac{2GM}{r}\right) dt^2 - \left(1 - \frac{2GM}{r}\right)^{-1} dr^2 - r^2 d\theta^2 - r^2 \sin^2\theta d\varphi^2, \tag{31}$$

let the frame $F_g$ be fixed with respect to the black-hole, and frame $F$ be moving with relative velocity $v$ between the frames $F$. (This corresponds to radial motion) We have, the Schwarszchild special relativistic factor,

$$\gamma_{Sch} = \frac{1}{\sqrt{1 - \left(1 - \frac{2GM}{r}\right)^{-2} v^2}}. \tag{32}$$

We note that for large values of $r$, $\gamma_{Sch} \to \gamma$, the Minkowski factor. This also happens for small values of Black hole mass $M$, as well as the gravitational constant $G$.

For Friedmann-Robertson-Walker (FRW) [4], with the line element,

$$ds^2 = dt^2 - R^2(t)\left(\frac{dr^2}{1 - kr^2} + r^2 d\theta^2 + r^2 \sin^2\theta d\varphi^2\right), \tag{33}$$

we have, we have the FRW special relativistic factor,

$$\gamma_{FRW} = \frac{1}{\sqrt{1 - \frac{R^2(t)v^2}{1 - kr^2}}}, \tag{34}$$

which again reduces to the Minkowski factor for large values of $r$.

For metrics with an off diagonal element $g_{10}$ equation for $\gamma_g$ turns out to be same as eq.(10). For Gödel universe [4], with the line element,

$$ds^2 = a^2\left(dt^2 - 2e^x dtdy - dx^2 + \frac{1}{2}e^{2x}dy^2 - dz^2\right), \tag{35}$$

special relativistic factor $\gamma_{Godel}$ is same as Minkowski factor $\gamma$, for 4-velocity $V^\mu=(1,v,0,0)$, which is relative motion along $x$ direction. However, for $V^\mu=(1,0,v,0)$, which represents relative motion along $y$ axis, the special relativistic factors gets modified,

$$\gamma_{Godel} = \frac{1}{\sqrt{1 - 2e^x v - \frac{1}{2}e^{2x}v^2}}. \tag{36}$$

Both frames F and F$_g$ are inertial and rotating with compass of inertia.

**4. New Horizons**

From eq.(12) it follows that $\gamma_g$ becomes imaginary for,

$$g_{\mu\nu}V^\mu V^\nu < 1, \tag{37}$$

signaling paradoxical tachyonic states, in the sense that, due to space-time curvature, mass is becoming imaginary, for velocities less than that of light. For instance in the case of Schwarszchild black hole $\gamma_{Sch}$ becomes imaginary for,

$$r > \frac{2GM}{1-v}. \tag{38}$$

For FRW universe, $\gamma_{FRW}$, becomes imaginary for,

$$r > \sqrt{\frac{1 - R^2(t)v^2}{k}}. \tag{39}$$

While for Gödel universe, $\gamma_{Godel}$, becomes imaginary (for $V^\mu=(1,0,v,0)$, i.e. relative motion $v$ along $y$ direction), for,

$$x > \ln\left(\frac{\sqrt{6}-2}{v}\right). \tag{40}$$

## 5. Further work

It will be interesting to investigate classical and quantum fields, with modified STR in a curved space-time setting. Quantum field theory replaces energy-momemtum relations of STR by relations between energy-momentum operators. On a curved space-time, modified STR energy-momentum relations would have to be implemented.

Horizons in various space-times, such as FRW, black-holes, Gödel universe, AdS universe would have to be redefined for various inertial frames, taking into account their velocities. It would also be interesting to investigate effect on horizons, for accelerating frames.

Closed Time-like Curves (CTCs) occur in a variety of situations in General relativity, and are a subject of an ongoing debate [5,6]. Some of them such as the worm hole CTC [7], and Gott CTC [8], contain velocity appearing in components of metric tensor, i.e., the CTC construction is based upon time dilation due to relative velocity. Effect of STR in a curved space-time setting, on these CTCs needs to be considered. Certain other CTCs, such as the Gödel universe CTC [4], the Kerr black hole CTC [9], the Tomimatsu-Sato solution's CTC [10], von-Stockum universe's CTC [11], are linked to rotation. Rotation in STR has also been a subject of extensive debate [12], to which discussions in this paper may contribute, via definition of $\gamma_g$.